\documentclass[showpacs,prl,preprintnumbers,amsmath,amssymb]{revtex4}

\usepackage{graphicx}
\usepackage{dcolumn}
\usepackage{bm}

\begin{document}

\title{General elastic interaction in nematic liquid crystals colloids}

\author{S. B. Chernyshuk $^{1,2}$,and B. I. Lev $^{2}$, }

\affiliation{$^{1}$, Institute of Physics, NAS Ukraine, Prospekt
Nauki 46, Kyiv 03650, Ukraine. $^{2}$, Bogolyubov Institute of
Theoretical Physics,NAS Ukraine,Metrologichna 14-b, Kyiv 03680,
Ukraine.}

\date{\today}

\begin{abstract}

The new free energy functional that describes general elastic
interaction between colloidal particles and nematic liquid
crystal has been proposed. It generalizes results of the paper
\cite{lupe} on the case of arbitrary orientation of colloidal
particles and is valid for arbitrary surface anchoring strength.
Formal analogies and differences between electric particles and
colloidal particles in LC are found. It is first time shown that
spur of the quadrupole moment tensor is different from zero
$Sp\hat{Q}_{\mu}\neq 0$ in the case of axial symmetry breaking of
the director field around colloidal particle in contrast to
electrostatics. New general presentation for the elastic
interaction between arbitrary colloids is obtained which contains new multipole terms connected with $Sp\hat{Q}_{\mu}\neq 0$ that are absent in the standard electrostatics.

\end{abstract}

\maketitle

Colloidal particles in liquid crystals (LC) have attracted a
great research interest during the last years. Anisotropic
properties of the host fluid - liquid crystal give rise to a new
class of colloidal anisotropic interactions that never occurs in
isotropic hosts. Liquid crystal colloidal system have much recent
interest as models for diverse phenomena in condense matter
physics. The anisotropic interactions result in different
structures of colloidal particles such as linear chains in
inverted nematic emulsions \cite{po1,po2}, 2D crystals \cite{Mus}
and 2D hexagonal structures at nematic-air interface
\cite{nych,R10}.

Study of anisotropic colloidal interactions has been made both
experimentally \cite{po2}-\cite{jap} and theoretically
\cite{lupe}-\cite{Gen}. The first theoretical approach was
developed in \cite{po1,lupe} with help of ansatz functions for
the director and using multiple expansion in the far field area.
Authors investigated spherical particles with hyperbolic hedgehog
and found dipole and quadrupole elastic interactions between such
particles. Another approach \cite{R5,lev} gave possibility to
find approximate solutions in terms of geometrical shape of
particles for the case of small anchoring strength and has
provided the way to connect the type of the interaction potential
with the local symmetry of the director field around particles
\cite{lev3}. The concept of coat has been introduced that
contains all topological defects inside and carries the symmetry
of the director and it enables to determine qualitatively the
type of the interaction potential. It was first shown there that
axial symmetry breaking leads to the origin of elastic charge and
coulomb-like interaction. But the coat is not quantitatively
exactly defined. In \cite{perg1,perg2} authors suggested approach
by fixing director field on the surface of the imaginary sphere.
There multiple expansion of the potential was obtained as well
but coefficients differ significantly from the results of
\cite{lupe}. For instance dipole-dipole interaction was found to
be three times weaker, and quadrupole five times weaker than in
\cite{lupe}. On the other hand authors of \cite{jap} recently
have measured experimentally both interactions and found that
results are in accordance with Lubensky at all prediction with
about $10\%$ accuracy. This allows to confirm rightness of the
main assumptions of \cite{lupe} for spherical particles though it
do not work for particles with broken axial symmetry. In this
paper we suggest to generalize that results for the case of
arbitrary particle's orientation. We propose new free energy
functional that describes interaction of colloidal particles with
director field and turns into functional of \cite{lupe} for axial
particles. It is valid for both strong anchoring case with the
presence of topological defects and for weak anchoring as well.
We show that axial symmetry breaking leads to the violation of $Sp\hat{Q}_{\mu}=0$ condition for spur of the quadrupole moment of particles in contrast to the classical electrostatics. This leads to the principal new multipole 
terms in elastic interaction connected with $Sp\hat{Q}_{\mu}$ which are absent in the
classical electrostatics. We show that the reason of such difference leads in nonlinearities of director field on close distances and in surface driven effects associated with anchoring.

In order to explain subsequent speculations we consider first the
case of homogeneous liquid crystal with uniform director
$\mathbf{n_{0}}$ and one particle immersed into it. Anchoring of
the liquid crystal with the surface of the particle produces deformations
of the director field around the particle so that director
$\mathbf{n(\mathbf{R})}$ varies from point to point. In the
one-constant approximation the total  free energy of the system:
particle plus liquid crystal has the form:
\begin{equation}
F=\frac{K}{2}\int dV \left[(div
\mathbf{n})^{2}+(rot\mathbf{n})^{2}\right]+\oint dS W(\mathbf{\nu }\cdot
\mathbf{n})^{2}\label{fto}
\end{equation}
where $W$ is anchoring strength coefficient, $\nu$ is the unit
normal vector, integration $\oint dS$ is carried out on the
surface of the particle, $K$ is the Frank elastic constant.

Far from the particle director field variations are small
$\mathbf{n(\mathbf{R})}\approx(n_{x},n_{y},1)$, $|n_{\mu}|\ll 1$
($\mu=x,y$) and bulk free energy has the form:
\begin{equation}
F_{b,linear}=\frac{K}{2}\int dV \left\{(\nabla n_{x})^{2}+(\nabla n_{y})^{2}\right\}\label{flin}
\end{equation}
which brings Euler-Lagrange equation of Laplace type:
\begin{equation}
\Delta n_{\mu}=0 \label{nmu}
\end{equation}
At large distances $R$ in general case it can be expanded in
multiples,
\begin{equation}
n_{\mu}(\mathbf{R})=\frac{q_{\mu}}{R}+\frac{\textbf{p}_{\mu}\textbf{u}}{R^{2}}+3\frac{\textbf{u}:\hat{Q}_{\mu}:\textbf{u}}{R^{3}}-\frac{Sp\hat{Q}_{\mu}}{R^{3}}+...
\label{nmufar}
\end{equation}
with $u_{\alpha}=R_{\alpha}/R$;
$\textbf{p}_{\mu}\textbf{u}=p_{\mu}^{\alpha}u_{\alpha}$,
$\textbf{u}:\hat{Q}_{\mu}:\textbf{u}=Q_{\mu}^{\alpha\beta
}u_{\alpha}u_{\beta}$, $Sp\hat{Q}_{\mu}=Q_{\mu}^{\alpha\alpha}$
and $\mu=(x,y)$. This is the most general expression for the
director field. It differs from usual electrodynamics with
$-Sp\hat{Q}_{\mu}/R^{3}$ term. We shall show below that it is not
zero when the axial symmetry breaking takes place. We note that the
multiple expression does not depend on the anchoring strength. It
is valid on far distances for any anchoring, weak and strong,
without topological defects or with them. Of course in order to
find multiple coefficients we need to solve the problem in the
near nonlinear area either with computer simulation or Ansatz
functions. Let's imagine that we have found all multiple
coefficients for the particular particle (for instance with help of
computer simulation): 20 different values, with 10 for every
$n_{\mu}$. So for every $\mu=(x,y)$ we have one elastic charge
$q_{\mu}$, three components of the vector dipole moment
$p_{\mu}^{\alpha}$ and six components of the quadrupole tensor
$\hat{Q}_{\mu}$ (it is symmetric
$Q_{\mu}^{\alpha\beta}=Q_{\mu}^{\beta \alpha}$, but
$Sp\hat{Q}_{\mu}\neq 0$ in general ). How particles will
elastically interact if we know all these 20 multiple
coefficients. We want to obtain analytical interaction potential.
Our task is to build effective harmonic free energy functional
which includes all multiple coefficients and nothing else. It
must be invariant and it's Euler-Lagrange equations for one
particle must have solution in the form of multiple expansion
(\ref{nmufar}). It is easy to check that such functional has the
following form:
\begin{equation}
F=K \int d^{3}x   \left\{ \frac{(\nabla n_{\mu})^{2}}{2} -4\pi q_{\mu}(\textbf{x})n_{\mu}-4\pi p_{\mu}^{\alpha}(\textbf{x})\partial_{\alpha}n_{\mu}-4\pi Q_{\mu}^{\alpha \beta}(\textbf{x})\partial_{\alpha}\partial_{\beta}n_{\mu} \right\}\label{fmain1}
\end{equation}
In the case of $N$ particles
$q_{\mu}(\textbf{x})=\sum_{i}q_{\mu,i}\delta(\textbf{x}-\textbf{x}_{i});
p_{\mu}^{\alpha}(\textbf{x})=\sum_{i}p_{\mu,i}^{\alpha}\delta(\textbf{x}-\textbf{x}_{i});
Q_{\mu}^{\alpha \beta}(\textbf{x})=\sum_{i}Q_{\mu,i}^{\alpha
\beta}\delta(\textbf{x}-\textbf{x}_{i})$ with $i$ being the number
of the particle. Euler-Lagrange equations for functional
(\ref{fmain1}) are Poisson like:
\begin{equation}
\Delta n_{\mu}=-4\pi q_{\mu}(\textbf{x})+4\pi [\partial_{\alpha}(p_{\mu}^{\alpha}(\textbf{x}))-\partial_{\alpha}\partial_{\beta}Q_{\mu}^{\alpha \beta}(\textbf{x}) ]\label{eqmain3}
\end{equation}
Substitution of this equation into (\ref{fmain1}) leads to the
following expression for the total energy of the system: particles
and liquid crystal:
\begin{equation}
F_{total}=-2\pi K \int d^{3}x   \left\{ q_{\mu}(\textbf{x})n_{\mu}^{ex}+(\textbf{p}_{\mu}(\textbf{x})\cdot\nabla) n_{\mu}^{ex}+(\hat{Q}_{\mu}(\textbf{x}):\nabla:\nabla) n_{\mu}^{ex} \right\}\label{fmain2}
\end{equation}
where $n_{\mu}^{ex}$ -exact solution of the (\ref{eqmain3}). For
unlimited space the solution of equation (\ref{eqmain3}) has
standard expression:
\begin{equation}
n_{\mu}(\textbf{x})=\int d^{3}\textbf{x}' \frac{1}{\left|\textbf{x}-\textbf{x}'\right|}\left[ q_{\mu}(\textbf{x}')-\partial_{\alpha}'(p_{\mu}^{\alpha}(\textbf{x}'))+\partial_{\alpha}'\partial_{\beta}'Q_{\mu}^{\alpha \beta}(\textbf{x}') \right] \label{solmain}
\end{equation}
For one particle we have
$q_{\mu}(\textbf{x})=q_{\mu}\delta(\textbf{x}),p_{\mu}^{\alpha}(\textbf{x})=p_{\mu}^{\alpha}\delta(\textbf{x}),Q_{\mu}^{\alpha
\beta}(\textbf{x})=Q_{\mu}^{\alpha \beta}\delta(\textbf{x})$
which leads us immediately to the general expression
(\ref{nmufar}). For axially symmetric particles $q_{\mu}=0$ and
the director take the form
$n_{x}=p\frac{x}{R^{3}}+6Q\frac{xz}{R^{5}}$,$n_{y}=p\frac{y}{R^{3}}+6Q\frac{yz}{R^{5}}$.
In Lubensky notation \cite{lupe} $Q=\frac{c}{3}$.

Actually in the construction of the functional (\ref{fmain1}) we
follow Lubensky ideology but with one significant difference. In
the paper \cite{lupe} authors tried to build invariants of
vector  $\textbf{P}$, tensor $C_{ij}$ and director vector
$\textbf{n}$, which has three components,not two. So obtained
invariants are of the type
 $-\textbf{P}\cdot\textbf{n}(\nabla \textbf{n}), (\nabla \textbf{n})\textbf{n}\cdot
\nabla(n_{i}C_{ij}n_{j})$ and include nonlinear terms of
$n_{\mu}$. In linear approximation authors find the following main
elastic functional:
\begin{equation}
F=K \int d^{3}x   \left\{ \frac{(\nabla n_{\mu})^{2}}{2} -4\pi P_{z}(\textbf{x})\partial_{\mu}n_{\mu}-4\pi C_{zz}(\textbf{x})\partial_{z}\partial_{\mu}n_{\mu} \right\}\label{fmainlub}
\end{equation}
This functional is axially symmetric, it correctly describes
axially symmetric particles. But in the case of axial symmetry
breaking it does not valid. In the case of axial symmetry breaking
we need to take into account coulomb like expression and to
introduce two different vectors $\textbf{p}_{x},\textbf{p}_{y}$
and two tensors $\hat{Q}_{x},\hat{Q}_{y}$ for each component
$\mu$. Using them we find new functional (\ref{fmain1}) which
describes system of any particles immersed into NLC. Let's show how
(\ref{fmain1}) proceeds to the (\ref{fmainlub}) for axially
symmetric particles.  In this case
$\textbf{p}_{x}=(p,0,0);\textbf{p}_{y}=(0,p,0)$ and quadrupole
tensors are of the form:
\begin{equation}
\hat{Q}_{x}=\left(\begin{array}{clrr}    0 & 0 & Q  \\   0  & 0 & 0 \\ Q & 0 & 0  \end{array}\right)
\hat{Q}_{y}=\left(\begin{array}{clrr}    0 & 0 & 0  \\   0  & 0 & Q \\ 0 & Q & 0  \end{array}\right).
\end{equation}
Then automatically (\ref{fmain1}) proceeds to the (\ref{fmainlub}) with
$P_{z}(\textbf{x})=p(\textbf{x});
C_{zz}(\textbf{x})=2Q(\textbf{x})$.

It is necessary to note that functional (\ref{fmain1}) defines
energy of the colloidal particle in the director deformations
created by any external sources. So if we have some deformed NLC
the particle will move in the way to minimize (\ref{fmain1}).

Director field from \textit{N} immersed particles will be superposition of
deformations (\ref{nmufar}) from all particles $
n_{\mu}(\textbf{x})=
\sum_{i}\frac{q_{\mu,i}}{\left|\textbf{x}-\textbf{x}_{i}\right|} +
\sum_{i}\frac{\textbf{p}_{\mu,i}(\textbf{x}-\textbf{x}_{i})}{\left|\textbf{x}-\textbf{x}_{i}\right|^{3}}+
\sum_{i}\frac{3(\textbf{x}-\textbf{x}_{i}):\hat{Q}_{\mu,i}:(\textbf{x}-\textbf{x}_{i})}{\left|\textbf{x}-\textbf{x}_{i}\right|^{5}}-
\sum_{i}\frac{Sp\hat{Q}_{\mu,i}}{\left|\textbf{x}-\textbf{x}_{i}\right|^{3}}\label{solmain}$.
Substituting this director into (\ref{fmain2}) and neglecting
with self energy divergent parts we come to the fact that total
energy of the system may be presented as the sum of pair
interactions: 

\begin{equation}
F_{total}=\sum_{i}\sum_{j>i}U^{ij}(\textbf{x}_{i}-\textbf{x}_{j})
\label{ftotpair}
\end{equation}

\begin{equation}
U^{ij}=U_{qq}+U_{qd}+U_{dd}+U_{qQ}+U_{dQ}+U_{QQ}
\label{fpair}
\end{equation}

\begin{equation}
U_{qq}=-4\pi K \frac{q_{\mu,i}q_{\mu,j}}{r_{ij}}
\label{fpair0}
\end{equation}

\begin{equation}
U_{qd}=-4\pi
K \frac{q_{\mu,i}(\textbf{p}_{\mu,j}\cdot
\textbf{u}_{ij})-q_{\mu,j}(\textbf{p}_{\mu,i}\cdot
\textbf{u}_{ij})}{r_{ij}^2}
\label{fpair1}
\end{equation}

\begin{equation}
U_{dd}=-4\pi K
\frac{(\textbf{p}_{\mu,i}\cdot
\textbf{p}_{\mu,j})-3(\textbf{p}_{\mu,i}\cdot
\textbf{u}_{ij})(\textbf{p}_{\mu,j}\cdot
\textbf{u}_{ij})}{r_{ij}^3}
\label{fpair2}
\end{equation}

\begin{equation}
U_{qQ}=-12\pi K  \left\{\frac{q_{\mu,i}(\textbf{u}_{ij}:\hat{Q}_{\mu,j}:\textbf{u}_{ij})+q_{\mu,j}(\textbf{u}_{ij}:\hat{Q}_{\mu,i}:\textbf{u}_{ij})}{r_{ij}^3}-
\frac{(q_{\mu,i}Sp\hat{Q}_{\mu,j}+q_{\mu,j}Sp\hat{Q}_{\mu,i}}{3r_{ij}^3})
\right\}
\label{fpair3}
\end{equation}

\begin{equation}
U_{dQ}=-12\pi K \left\{\frac{(\textbf{p}_{\mu,i}\cdot \textbf{u}_{ij})Sp\hat{Q}_{\mu,j}-(\textbf{p}_{\mu,j}\cdot \textbf{u}_{ij})Sp\hat{Q}_{\mu,i} }{r_{ij}^4}+2\frac{(\textbf{p}_{\mu,i}:\hat{Q}_{\mu,j}:\textbf{u}_{ij}-\textbf{p}_{\mu,j}:\hat{Q}_{\mu,i}:\textbf{u}_{ij})}{r_{ij}^4}\right\}-
\label{fpair4}
\end{equation}

\[
-60\pi K \frac{(\textbf{u}_{ij}:\hat{Q}_{\mu,i}:\textbf{u}_{ij})(\textbf{p}_{\mu,j}\cdot \textbf{u}_{ij})-(\textbf{u}_{ij}:\hat{Q}_{\mu,j}:\textbf{u}_{ij})(\textbf{p}_{\mu,i}\cdot \textbf{u}_{ij})}{r_{ij}^4}
\]

\begin{equation}
U_{QQ}=-12\pi K \left\{\frac{2\hat{Q}_{\mu,i}::\hat{Q}_{\mu,j}-20\textbf{u}_{ij}:\hat{Q}_{\mu,i}:\hat{Q}_{\mu,j}:\textbf{u}_{ij}+35(\textbf{u}_{ij}:\hat{Q}_{\mu,i}:\textbf{u}_{ij})(\textbf{u}_{ij}:\hat{Q}_{\mu,j}:\textbf{u}_{ij})}{r_{ij}^5}\right\}-
\label{fpair5}
\end{equation}

\[
-12\pi K \frac{Sp\hat{Q}_{\mu,i}Sp\hat{Q}_{\mu,j}}{r_{ij}^5}
\]

where
$\textbf{u}_{ij}=\frac{\textbf{x}_{i}-\textbf{x}_{j}}{\left|\textbf{x}_{i}-\textbf{x}_{j}
\right|}$ and $r_{ij}=\left|\textbf{x}_{i}-\textbf{x}_{j}\right|$.
These expressions are the most general elastic pair interaction
potentials between ordinary particles immersed into NLC. They are
valid for particles with topological defects and without them,
for any type and strength of the anchoring coefficient. Coulomb-like elastic attraction and repulsion between inclined ellipsoidal particles was found in \cite{miha},\cite{new} with help of computer simulation. We consider that Coulomb-like elastic interaction takes place between any non-symmetrical particles with broken axial symmetry (for instance non-symmetrical triangles or cubes with different anchoring on sides) even in the case of zero torque applied to the particles.

If particles are the same then $U_{qd}\equiv 0, U_{dQ}\equiv 0$. We
see that formulas (\ref{fpair3})-(\ref{fpair5}) contains new
terms connected with $Sp\hat{Q}_{\mu}$ which are absent in the
classical electrostatics. 
Let's analize this result in comparison with electrostatics. In electrostatics there is the charge density distribution $\rho(\textbf{x})$ in the Poisson equation $\Delta\varphi=-4\pi\rho$ and linear electrostatics works \textit{inside} this density. So \textit{inside} $\rho(\textbf{x})$ there is superposition principle. But in nematostatics there is \textit{principle} difference on small distances - the superposition principle does not work because of nonlinear deformations. So in case of strong anchoring we consider that the notion of elastic charge density is unreasonable. That's why in nematostatics we introduce the concept of elastic charge $q_{\mu}$, dipole moment $\textbf{p}_{\mu}$, quadrupole moment $Q_{\mu}^{\alpha\beta}$ not as compound expressions like in electrostatics (full charge is  equal the sum of compound charges $q_{total}=\sum_{i}q_{i}$, dipole moment as $\textbf{P}=\sum_{i}q_{i}\textbf{r}_{i}$ and so on ), but as primary quantities for each particle, the quantities which are just the linear asymptotics of exact \textit{nonlinear solutions}. 

As well in nemato-statics surface effects can make $Sp\hat{Q}_{\mu}\neq 0$. Let's show this
on the example of small particles with weak anchoring without
topological defects. Then deformations are small anywhere
$\mathbf{n(\mathbf{R})}\approx(n_{x},n_{y},1)$, $|n_{\mu}|\ll 1$
($\mu=x,y$) and surface energy takes the form  $\oint ds
W(\mathbf{\nu }\cdot \mathbf{n})^{2}\approx const+2W\oint ds
\nu_{\mu}\nu_{3}n_{\mu}(\textbf{s}) $. Taylor expansion of the
director concerning center of the particle brings : $
n_{\mu}(s)=n_{\mu}(x_{0})+(l\cdot\nabla)n_{\mu}(x_{0})+\frac{1}{2}(l\cdot\nabla)^{2}n_{\mu}(x_{0})
\label{nro}$ where $\textbf{l}$ is the vector from the center
$\textbf{x}_{0}$ to the point on the surface
$\textbf{s}=\textbf{x}_{0}+\textbf{l}$. Then we automatically
obtain that free energy (\ref{fto}) takes the form (\ref{fmain1})
with such multiple coefficients: $\tilde{q}_{\mu}=-\frac{W}{2\pi
K}\oint ds \nu_{3}\nu_{\mu} \label{qweak}$,
$\tilde{\textbf{p}}_{\mu}=-\frac{W}{2\pi K}\oint ds
\nu_{3}\nu_{\mu}\textbf{l}(s)$,
$\tilde{Q}_{\mu}^{\alpha\beta}=-\frac{W}{4\pi K}\oint ds
\nu_{3}\nu_{\mu}l_{\alpha}(s)l_{\beta}(s)$. These quantities are
of course approximate as Taylor extension is not exact but we
consider that qualitative estimates may be correctly performed.
Let's note that substituting them into $U_{ij}$  we naturally
come to the results of papers \cite{lev,lev3}.

So we may clearly see that $Sp\tilde{Q}_{\mu}=-\frac{W}{4\pi
K}\oint ds \nu_{3}\nu_{\mu}l^{2}(s)\approx \left\langle
l^{2}\right\rangle \tilde{q}_{\mu} \neq 0(s)$ if
$\tilde{q}_{\mu}\neq 0$, so that $Sp\tilde{Q}_{\mu}$ and
$\tilde{q}_{\mu}$ are approximately proportional to each other (
here $\left\langle l^{2}\right\rangle$ is average square of the
distance from the surface to the center of the particle). As the
elastic charge occurs when axial symmetry is broken \cite{lev3}
we come to the conclusion that in that case as well
$Sp\tilde{Q}_{\mu}\neq 0$. In the paper \cite{new} authors have
found Coulomb-like elastic interaction between two inclined
ellipsoidal particles using computer simulation. As well there
analytical expression for the elastic charge for the case of
small anchoring was obtained using
$\tilde{q}_{\mu}=-\frac{W}{2\pi K}\oint ds \nu_{3}\nu_{\mu}$.
Then it was found that $ \tilde{q}_{x}=-\frac{W}{2\pi
K}F(b,\varepsilon)\cos\psi \sin\theta \cos \theta$,
$\tilde{q}_{y}=-\frac{W}{2\pi K}F(b,\varepsilon)\sin\psi
\sin\theta \cos \theta$, where $\theta$ is the angle between long
axis and $ \textbf{n}_{0}$, $\psi$ is the angle between the
horizontal projection of the particle and direction of axis $x$,
$b$ is the length of the long axis and $\varepsilon$ is the
eccentricity $\varepsilon= \sqrt{1-c^{2}/b^{2}}$ with $c$ being
the short axis of the ellipsoid and:
\begin{equation}
F(b,\varepsilon)=\pi
b^{2}\sqrt{1-\varepsilon^{2}}\left\{\frac{\sqrt{1-\varepsilon^{2}}}{\varepsilon^{2}}(2\varepsilon^{2}-3)+
\frac{\arcsin \varepsilon
}{\varepsilon^{3}}(3-4\varepsilon^{4})\right\}. \label{fb}
\end{equation}
In the case of small eccentricities it has the asymptotic
expansion $F(b,\varepsilon)_{\varepsilon\Rightarrow
0}=-\frac{10}{6}\pi b^{2}\varepsilon^{2}+o(\varepsilon^{2}) $.
Here we come to an interesting conclusion, that the elastic
charge goes to zero as square of the eccentricity
$\tilde{q}_{\mu}\propto \varepsilon^{2}$. Then we may estimate
$Sp\tilde{Q}_{\mu}\approx \tilde{q}_{\mu} \left\langle
l^{2}\right\rangle$, where $\left\langle l^2\right\rangle$ is
something average between $b^{2}$ and $c^{2}$.

In this paper the new free energy functional which describes
elastic interaction between arbitrary colloidal particles and NLC
has been proposed. It contains multiple coefficients of each
colloidal particle and proceeds to Lubensky functional for
axially symmetric particles. This approach enables to find
elastic interaction energy between emersed particles in the most
general case for arbitrary anchoring strength on the surface. The
procedure to find interaction energy consists of two steps -
first, multiple coefficients for director field are found using
different methods like computer simulation, variational ansatz
and so on, second - these coefficients are used in the obtained
analytical expressions for elastic potentials. It was first time
shown that spur of the quadrupole moment tensor is different from
zero $Sp\hat{Q}_{\mu}\neq 0$ in the case of axial symmetry
breaking of the director field around colloidal particle in
contrast to classical electrostatics.

\end{document}